# Inverse Low Gain Avalanche Detector (iLGAD) Periphery Design for X-Ray Applications

A. Doblas[a,*], D. Flores[a], S. Hidalgo[a], N. Moffat[a], G. Pellegrini[a],
D. Quirion[a], J. Villegas[a], D. Maneuski[b], M. Ruat[c], P. Fajardo[c]

[a]Centro Nacional de Microelectronica, IMB-CNM-CSIC, Barcelona, Spain
[b]Physics and Astronomy Department, The University of Glasgow, Glasgow, UK
[c]European Synchrotron Radiation Facility, ESRF, Grenoble, France

**Abstract**

LGAD technology is established within the field of particle physics, as the baseline technology for the timing detectors of both the ATLAS and CMS upgrades at the HL-LHC. Pixelated LGADs have been proposed for the High Granularity Timing Detector (HGTD) and for the Endcap Timing Layer (ETL) of the ATLAS and CMS experiments, respectively. The drawback of segmenting an LGAD is the non-gain area between pixels and the consequent reduction in the fill factor. In this sense, inverse LGAD (iLGAD) technology has been proposed by IMB-CNM to enhance the fill factor and to reach excellent tracking capabilities. In this work, we explore the use of iLGAD sensors for X-Ray applications by developing a new generation of iLGADs. The periphery of the first iLGAD generation is optimized by means of TCAD tools, making them suitable for X-Ray irradiations thanks to the double side optimization. The fabricated iLGAD sensors exhibit good electrical performances before and after an X-Ray irradiation. The second iLGAD generation is able to withstand the same voltage, as contrary to the first iLGAD generation after irradiation.

*Keywords:* Radiation-hard detectors; Fast detectors; X-ray detectors; LGAD; Silicon; Low Energy X-ray

*Corresponding author

*Email address:* albertdoblas@gmail.com (A. Doblas)

## 1. Introduction

Low Gain Avalanche Detectors (LGADs) have been widely studied during the last years by the high energy physics community [1] - [5]. Based on the Avalanche Photodiode (APD) concept, the LGAD exhibits an intrinsic multiplication in the linear region, previous to the Geiger avalanche mode. A p-doped layer, which is called the multiplication layer, is diffused underneath the high n-doped layer, creating a high electric field region. The doping level of the multiplication layer is selected to reach a moderate gain (10-30), unlike the APD structure. This gain value allows the detector to widen its operating voltage range and maintain a low noise, reducing the effect of the noise associated with the multiplication mechanism. Charge generated in the detector by an incident particle is directly related to its energy. Therefore, the energy of a charged particle going through a detector can be determined by monitoring the generated transient current.

In this sense, the large area detectors must be segmented in order to establish the incident point of the particle and to enhance the global precision of the detection measurements. Strip and pixel detectors are used as tracking detectors, where the multiplication region (n+/p) is segmented, which can collect the charge of the incident particle. However, there is an area between the pixels/strips where the charge does not undergo multiplication (dead area). The fill factor is defined as the ratio of the active area (multiplication area) to the total sensor area (multiplication plus dead area). In order to reach the maximum fill factor, IMB-CNM has developed the concept of the Inverse Low Gain Avalanche Detector (iLGAD) structure, where the segmentation is performed at the p+ electrode of the detector ohmic side and the multiplication region is a large, uniform area diffused along the n+ electrode side [6] - [7].

In the iLGAD structure, the charge generated by the incident particle will be multiplied independently of the hit position. Figure 1 shows a comparison between LGAD and iLGAD structures. Unlike LGADs, the charge collection in the iLGAD is made by holes, which have a lower mobility than electrons. Nevertheless, iLGAD shows time resolutions in the range of 20 ps, which is a similar value to that obtained in LGADs [8]. Due to their 100% fill factor iLGADs are excellent tracking detectors. Figure 2 shows the result of a test beam in strip LGAD and iLGAD detectors. Two peaks are observed in the LGAD, corresponding to the non-gain (around 24 ke) and the gain regions (77 ke), while in the iLGAD only one peak is present (75 ke) due to the 100 % fill factor.

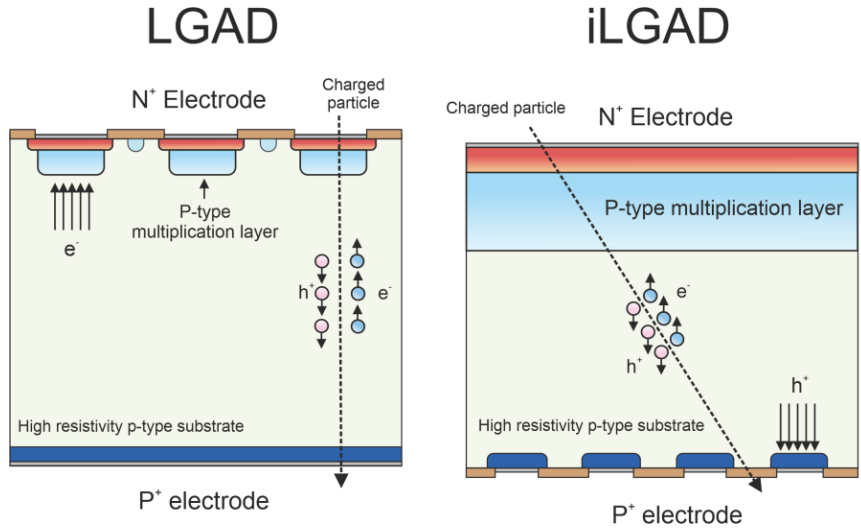

Figure 1: Comparison between LGAD and iLGAD structures. In the LGAD, the multiplication region is segmented on the n+ electrode side, collecting electrons. In the iLGAD, p+ electrode is segmented on the ohmic side, collecting holes. A 100% fill factor is achieved with the iLGAD. The sketch is not to scale.

Due to the intrinsic multiplication, low readout noise and 100 % fill factor, iLGAD technology is suitable for X-Ray detection, where thick detectors are required to collect all the charge generated by X-Rays. Lambert-Beer law states that the intensity of a photon decreases exponentially with the thickness of the material and depends on the incident photon energy [9]. A standard 300 $\mu$m silicon detector is able to absorb 13 keV X-Rays, but Silicon sensors are limited up to 20 keV, as the thickness of the active substrate beyond this energy must be > 1 mm. In this sense, for high-energy X-Rays, other semiconductor materials such as, CdTe and GaAs are under investigation due to their higher attenuation coefficient [10] - [11].

The main challenge of X-Ray detectors is the radiation damage generated at the Si-*SiO$_2$* interface. When a device is irradiated electron-holes pairs are created within the interface, as the electrons have a high mobility in silicon they are rapidly collected. However, as holes have a lower mobility, a cloud of holes is created inside the *SiO$_2$*. This cloud attracts electrons and an n-doped layer is created underneath the oxide. As a consequence, the electrical performance of the detector is significantly reduced due to the X-Ray irradiations. In this work, the design, fabrication at the IMB-CNM labs and characterisation of a suitable periphery region of an iLGAD for synchrotron applications is described.

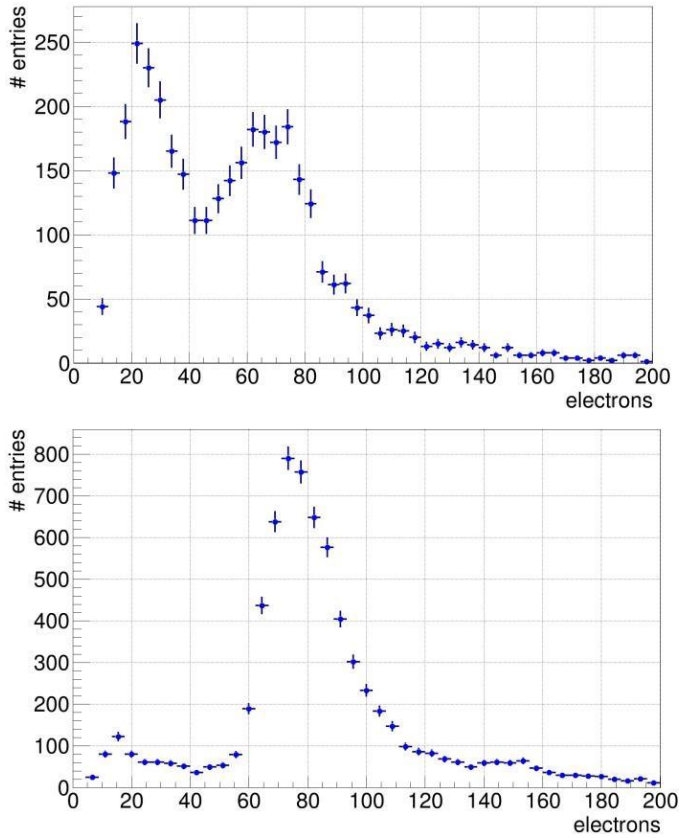

Figure 2: Charge distribution measured during a test beam for one strip LGAD and one strip iLGAD. The iLGAD shows a uniform gain. Graph taken from [8].

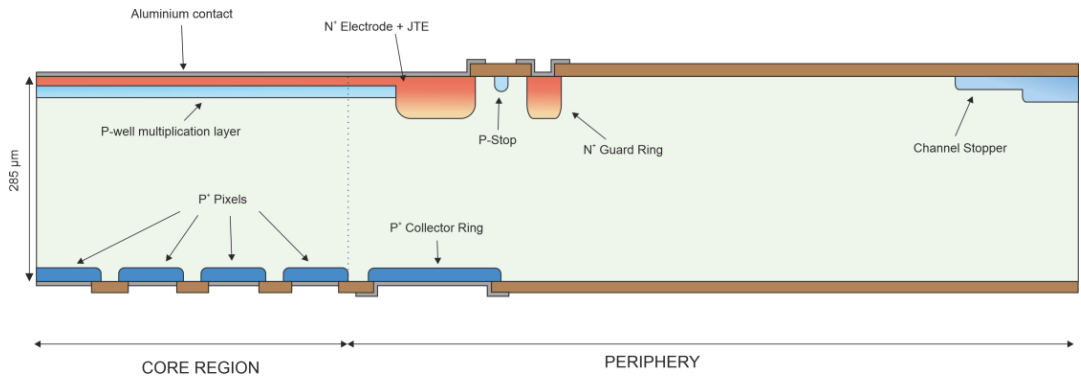

Figure 3: First iLGAD generation (iLG1) fabricated at IMB-CNM, not optimized for XRay irradiations, with a JTE and channel stopper. In the ohmic side, a collector ring is diffused in the periphery.

The periphery optimization is carried out with the aid of Technological Computer Aided Design (TCAD) tools. The new iLGAD design supports the X-Ray irradiations, maintaining the electrical performances of the detector. Furthermore, irradiation measurements have been carried out to determine the performance of the new iLGAD detector.

## 2. Optimization of the iLGAD periphery

As the multiplication region is the same for LGAD and iLGAD structures, only the periphery of the detector has to be optimized to make it suitable for synchrotron applications. The first iLGAD prototypes fabricated at IMBCNM (first iLGAD generation, iLG1) were not designed assuming a high concentration of charges at the Si-$SiO_2$ interface, since those devices were not used as X-Ray detectors. The periphery was initially optimized with a junction termination extension (JTE) at the edge of the p-well multiplication, introducing a channel stopper at the edge of the device and using high boron doped diffusions as collector rings at the ohmic side. An n-type collector ring is diffused at the multiplication side with a distance between the channel stopper and the n-type collector ring sufficiently large to avoid punch-through between both diffusions. Figure 3 shows a schematic view of the final iLG1 structure.

In order to simulate the damage created in the detector during X-Ray irradiations, a certain oxide charge density ($N_{ox}$) at the Si-$SiO_2$ interface has to be considered with a reference value of $N_{ox}$ = $10^{12}$ $cm^{-2}$, which corresponds to Φ = 10 MRad [16]. Figure 4 shows the effect of the charge density on the device breakdown voltage, where a strong reduction of the breakdown voltage is observed when the charge density increases. Thus, the effectiveness of the edge termination is drastically reduced in a harsh X-Ray radiation environment. It can be concluded that electrons are accumulated in the oxide-silicon interface and the resulting n-doped layer lowers the electrical performance of the device since it creates a high electric field at the edge of the channel stopper (p-type diffusion). In this situation, the peak electric field is higher than the electric field value created at the n+/p junction, leading to a premature breakdown.

In order to enhance the robustness of the device, this peak has to be reduced. The proposed solution is the addition of five floating rings with their respective p-stop diffusions between them, including two p-stops between the last floating ring and the channel stopper. A 50 V increase of the breakdown voltage is obtained using this strategy, which is less than expected. In order to further improve the breakdown voltage capability, the ohmic side of the device has also been optimized. Considering that p-type pixels and the collector ring at the ohmic side are highly-doped diffusions, the electric field created due to the conductive layer is higher than the main junction peak. The field plate contact in the collector ring is introduced to move the electric field peak towards the oxide, increasing the voltage breakdown up to 300 V. Using the same approach as in the multiplication side of the device, the ohmic side includes p-type floating rings and a high doped channel stopper at the edge. In this case, n-stop diffusions are not needed since the surface is highly n-doped. Figure 5 shows the electric field distribution along the length of the sensor, where the high peak electric field in the curvature of the diffusions can be clearly seen. The multi-ring strategy is the way to lower the peaks and to increase the breakdown voltage up to the 500 V range, as shown in figure 6.

As a conclusion, the breakdown voltage is almost four times in a harsh X-Ray environment, widening the operation voltage regime. Figure 7 shows the final design proposed for the iLGAD structure.

## 3. Fabrication of the prototypes

The main purpose of the iLGAD structure optimization is to make it suitable for X-Ray detection applications. In this sense, various pixelated sensor layouts have been included in the mask to connect to state-of-the-art readout ASICs for X-ray detectors [12], [13]. It enabled to check the impact of geometrical parameters on the sensor performance.

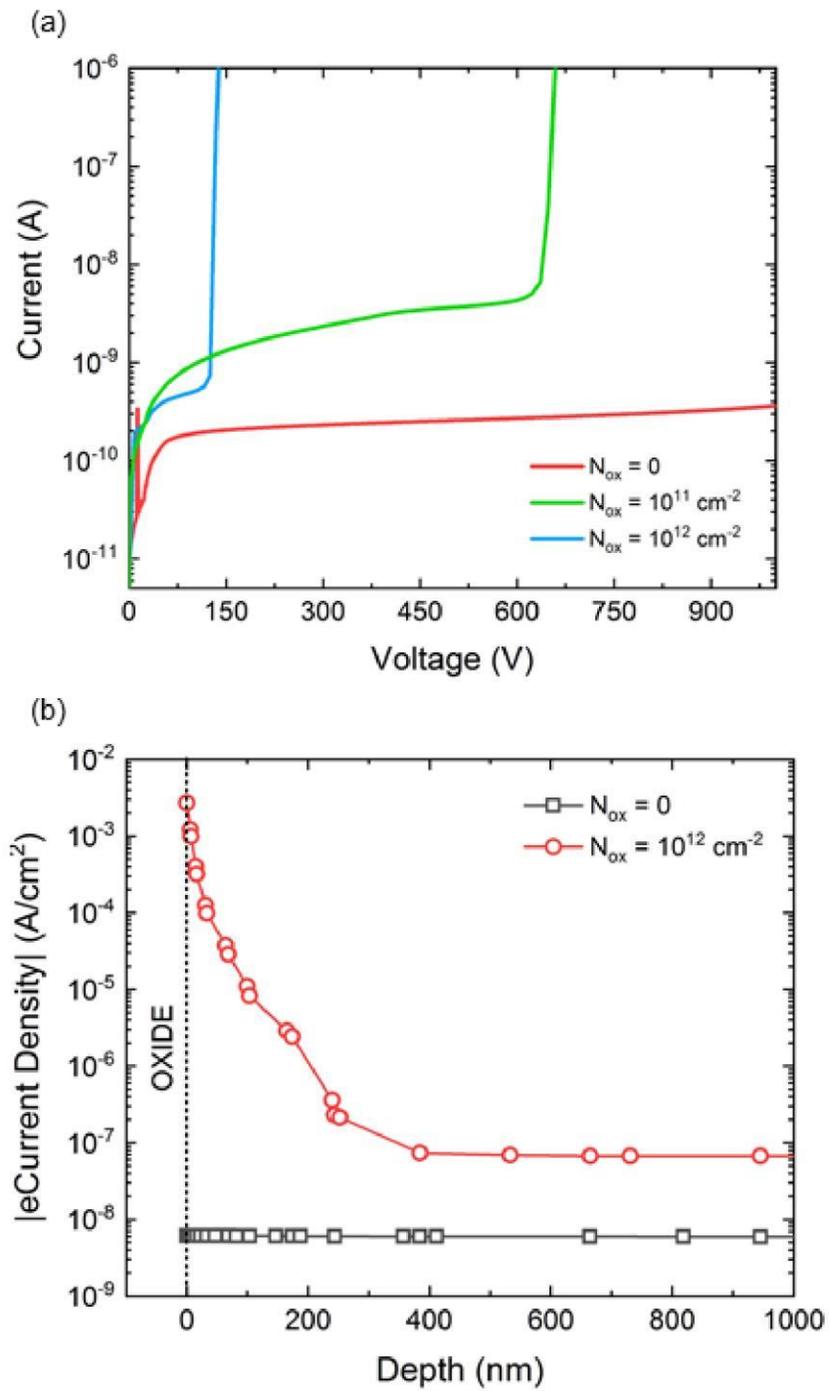

Figure 4: I-V simulation for an iLGAD with different $N_{ox}$ concentrations. Breakdown voltage is reduced by increasing $N_{ox}$.

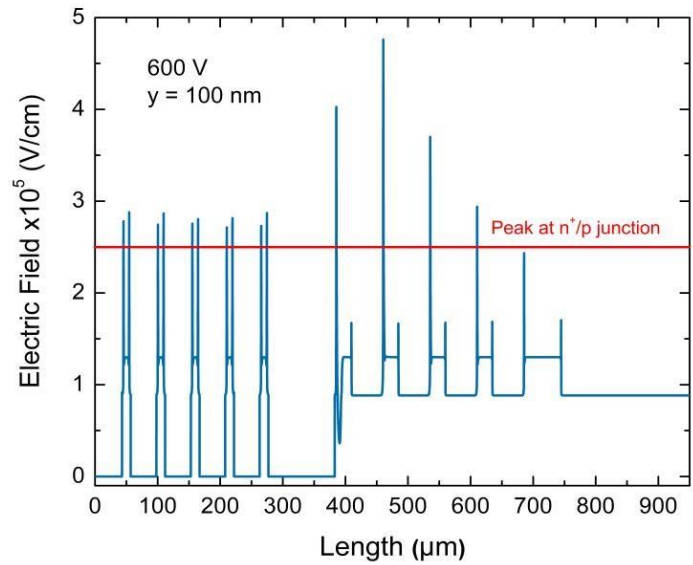

Figure 5: Electric field across the length of the iLGAD. Peaks at the periphery are higher than the maximum peak at the main junction. The conductive electron layer created by the X-Rays causes these peaks.

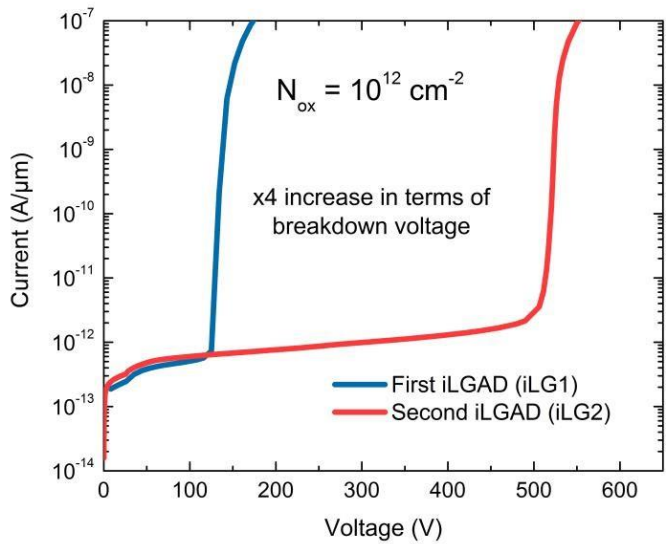

Figure 6: I-V simulation of iLGAD structures with $N_{ox}$ = $10^{12}\,cm^{-2}$. The initial and the optimized designs are included to show the significant increase of the breakdown voltage (almost four times).

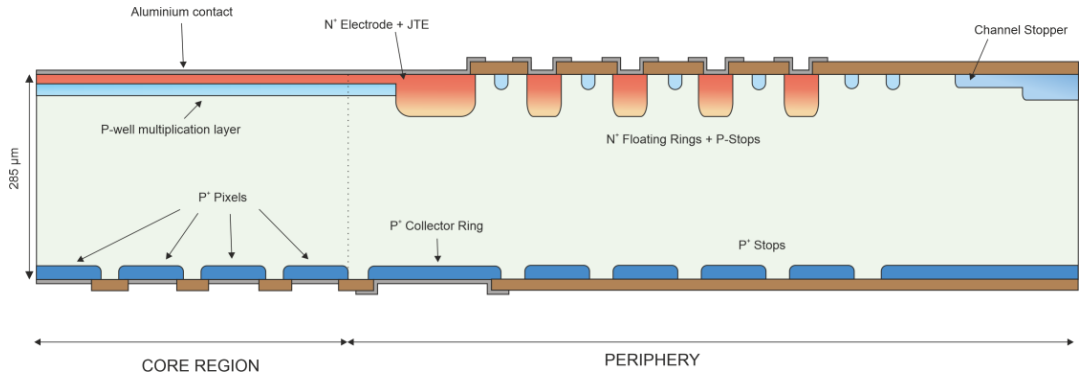

Figure 7: Schematic cross-section of the final periphery design including four n-type floating rings and p-stops at the multiplication side. At the ohmic side, we added four high doped p-stops and a field plate at the collector ring, in order to move the high electric field peak towards the oxide.

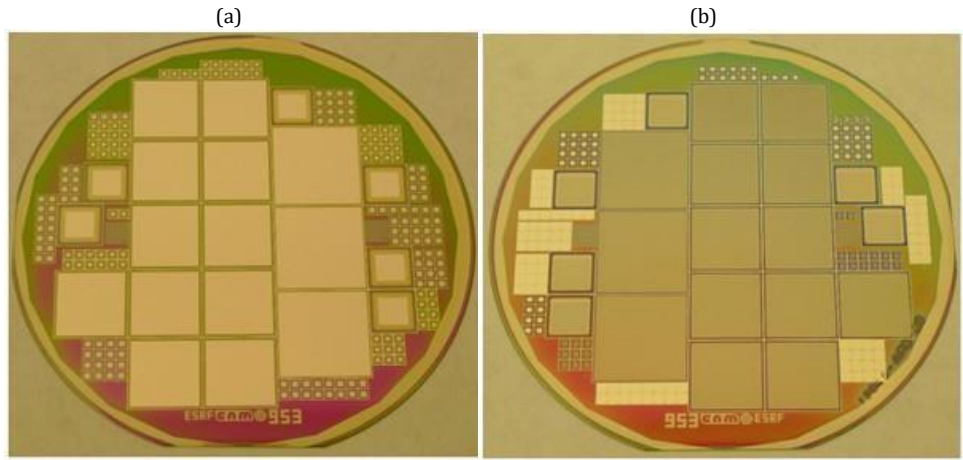

Figure 8: (a) Images of the multiplication side and (b) ohmic side images of the fabricated iLGAD wafers.

Pixelated sensors have 256 x 256 pixels with pixel pitch of either 55 $\mu$m or 75 $\mu$m. Additional detectors have also been designed in the mask set, including strip detectors, to test their performances and the yield of large area devices. Special emphasis is addressed in the correlation of the gain and voltage capability with the multiplication layer parameters. Moreover, small pixelated and pad-like detectors with the same periphery as Medipix3 have been included to test their resistance during X-Ray irradiations and metal-oxide-semiconductor (MOS) capacitors to study the oxide charges.

Finally, LGAD and PiN diodes have been included to control the technological parameters of the process technology such as gain, breakdown voltage and full depletion voltage ($V_{FD}$. Test detectors have been fabricated on 4-inch high resistivity (> 1 kΩ · cm) 285 $\mu$m thick p-type silicon wafers. Figures 8a and 8b show images of the processed wafers in the multiplication side (a) and the ohmic side (b), corresponding to the second iLGAD generation (iLG2) fabricated at the IMB-CNM labs.

The parameters used for the multiplication region are the same as for standard LGAD detectors since the X-Ray irradiation does not affect the performance of the gain layer, contrary to bulk damage caused by proton or a neutron irradiation. Medium dose and energy values of the multiplication implant have been selected for the fabrication process.

## 4. Electrical Characterisation

iLGAD prototypes have been tested to study their electrical performance. Figure 9 shows the main characteristics of these sensors. Figure 9a shows a 1D C-V simulation of an LGAD for different boron doses. As expected, depletion of the gain layer increases with boron dose. Therefore, gain is also increasing with boron dose, as one can observe in figure 9b. We expect to have a linear gain between 5-10. As already mentioned, non-pixelated sensors have been added to the mask set in order to test the technology. Figure 9c shows a C-V measurement of a pad-like iLGAD at room temperature where the gain layer depletion voltage ($V_{GL}$) is reached at 38 V and the full depletion at 70 V. The I-V measurement is shown in figure 9d, showing a leakage current in the range of 10 nA and a breakdown voltage of 450 V, lower than expected. Nevertheless, the operability range for these sensors is around 300 V.

A metal-oxide-semiconductor (MOS) capacitor is also characterized to extract the oxide charge of an un-irradiated sensor. Figure 9e shows the C-V measurement of the MOS capacitor with a flat-band voltage of -6.58 V. The calculated oxide charge [14] is $Q_{ox} \approx 10^{11}$ $cm^{-2}$, which is the typical value produced after a thermal wet oxidation process. The most critical parameter in LGADs is the gain, which is determined by using the transient current technique (TCT) [15]. For this purpose, a window is opened in the aluminium of the multiplication side, in order to allow an infrared (IR) laser to pass through. This measurement also performed on a pin diode in order to have a comparison for the gain calculation. Figure 9f shows the gain of the iLGAD, which is calculated by dividing the collected charges of each detector. A linear gain from 12 to 24 is measured in the 70-360 V range, which is higher than the expected. In conclusion, the doping concentration of the gain layer is slightly higher than the expected, resulting in a higher gain and a lower breakdown voltage. This can be caused by a non-uniformity in the thermal process after the multiplication implantation. Nevertheless, the detector is operative up to 350 V showing a medium linear gain, which is the objective of these technology.

## 5. Irradiated samples

In order to test the endurance of the new radiation-resistant periphery some detectors have been irradiated with X-rays at the X-ray Irradiation facility at the Glasgow Laboratory for Advanced Detector Development (GLADD). The detectors tested were: a pixelated iLGAD with an active area of 1x1 mm$^2$ from the second iLGAD generation and a strip iLGAD (8x8 mm$^2$) from the first iLGAD. Both detectors were irradiated at a fluence of $\Phi$ = 10 MRad. Figure 10 shows the current density of both irradiated and unirradiated detectors, measured in identical conditions.

The probes of the probe station are placed in contact the multiplication region, while the segmented side is in contact with the chuck (Ground). Therefore, the collector ring of the iLGAD corresponding to the iLG2 is not functioning, leading to a higher current density in this device. Unfortunately, problems when contacting the pixels have been encountered, as observed in the unirradiated measurements (probes are not properly functioning until 125 V). The detectors have been irradiated at $\Phi$ = 10 MRad, with a dose rate of 1.8 MRad/h. The iLG2 detector shows a slight increase in the leakage current, but the breakdown voltage capability is not affected.

On the other hand, for the iLG1 detector, the breakdown voltage is reduced to 75 V, making it of no use for X-Ray applications. Therefore, the irradiation measurements have demonstrated that the sensor has been successfully optimized for X-Ray applications, as the performance of the iLG2 detectors are much better than those of the first generation. Nevertheless, this statement should be corroborated with a thorough characterization to determine the robustness of the structure to X-ray irradiation.

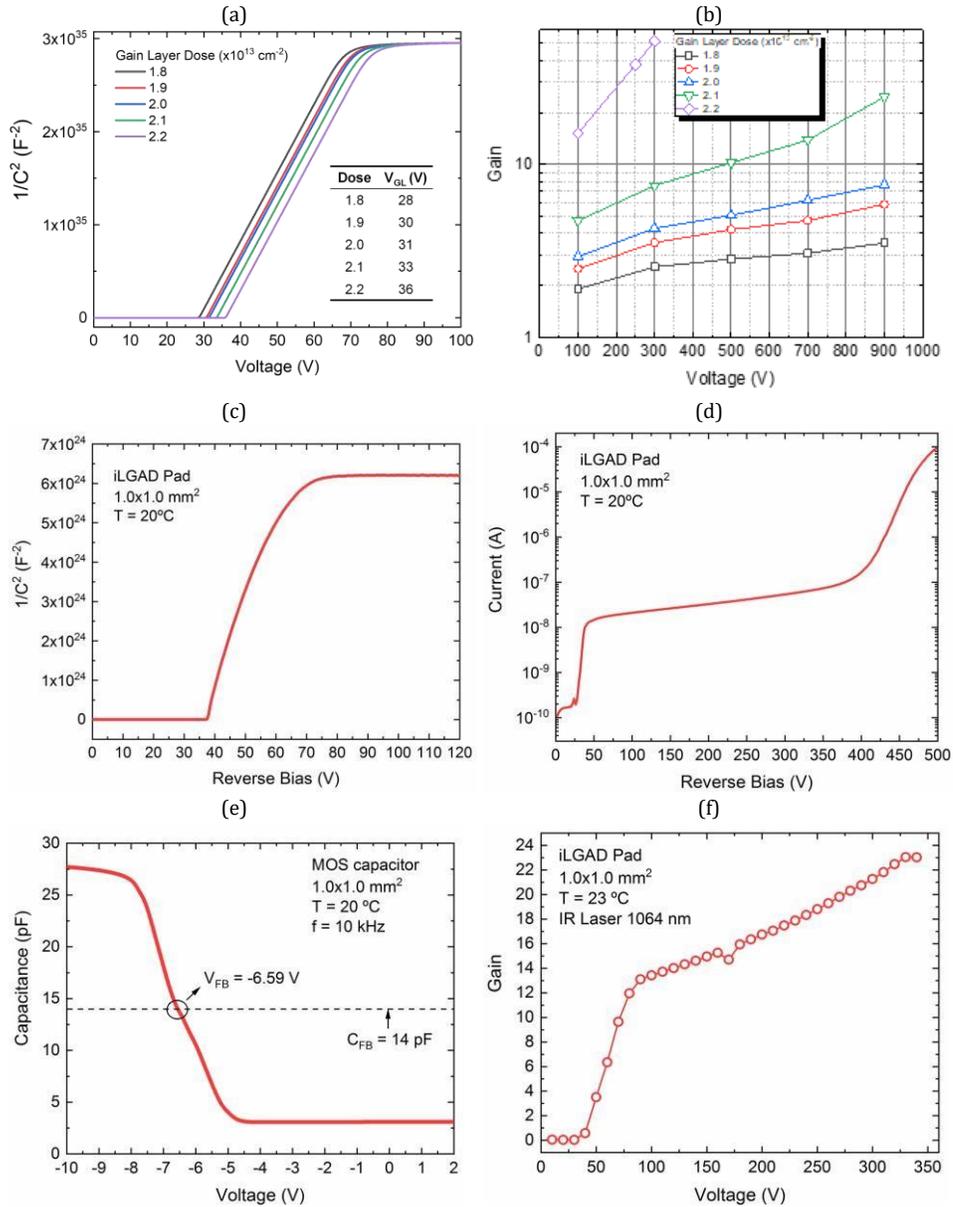

Figure 9: (a) C-V simulation for different boron doses. Gain layer is increasing with boron dose. (b) Gain simulations for each boron dose. As expected, gain is increasing with boron dose. (c) C-V measurement. Gain layer depletion is reached at 38 V. Full depletion voltage is 70 V. (d) I-V measurement. Leakage current of 10 nA until the breakdown, which is achieved over 400 V. (e) C-V measurement of a MOS capacitor. Flat-band voltage is graphically calculated using the flat-band capacitance value. (f) TCT measurement with an IR laser. Gain is calculated by diving the collected charge of an iLGAD by the charge of a pin diode. A linear gain between 12-24 is observed between 70-350 V.

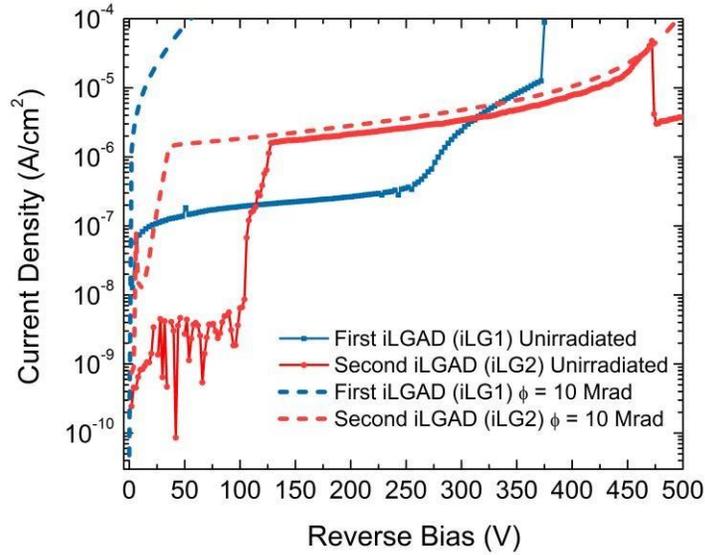

Figure 10: Current density of unirradiated and Φ = 10 MRad irradiated iLGAD detectors from the iLG1 and iLG2 runs. There is a difference in terms of breakdown voltage between both technologies. The iLG2 periphery is able to support a higher voltage in a harsh radiation environment.

## 6. Conclusion

Design and development of an iLGAD sensor for X-Ray applications is described in this paper. The new structure provides a 100% fill factor, in contrast with the LGAD technology that includes a dead zone between pixels. In order to use this structure for X-Ray detection, a suitable periphery has been designed by means of TCAD tools. In this sense, the limitations of the first generation against X-Ray irradiations are explained and the optimized periphery to achieve a higher operability voltage is proposed. Finally, the ohmic side of the detector has also been optimized to overcome the problems with the high electric field peaks at the curvature of the p+ diffusions. The detectors have been fabricated at the IMB-CNM clean room and electrically characterized.

Pad like iLGADs show a lower breakdown voltage and a higher leakage current than expected. This performance is related with the gain, leading to a gain in the range of 12-24, which is slightly higher than the gain obtained by simulation. C-V measurements have been performed to obtain the $V_{GL}$ and $V_{FD}$ of the detectors, in order to establish the range of operation. Moreover, CV measurements have been performed on MOS capacitors to extract the oxide charge of a non-irradiated sensor, which shows the expected result of $Q_{ox} \approx 10^{11}\ cm^{-2}$. Finally, some of the fabricated detectors have been irradiated in order to test the new periphery design and a comparison of the I-V behaviour between iLGADs of the first and second generation is reported.

The optimized design is able to withstand the same voltage before and after irradiation, while the sensor corresponding to the first iLGAD generation cannot be used at a voltage $> V_{FD}$, making it unusable for X-Ray applications.

**Credit authorship contribution statement**

**A. Doblas:** Formal analysis, Investigation, Data curation, Validation, Writing - original draft, Writing - review & editing. **D. Flores:** Conceptualization, Methodology, Validation, Resources, Data curation, Writing - review & editing, Work Supervision. **S. Hidalgo:** Conceptualization, Methodology, Validation, Resources, Data curation, Writing - review & editing, Work Supervision, Project administration, Funding acquisition. **N. Moffat:** Investigation, Writing - review & editing. **G. Pellegrini:** Conceptualization, Methodology, Validation, Resources, Funding acquisition. **D. Quirion:** Investigation. **J. Villegas:** Investigation, Data curation. **D. Maneuski:** Detectors X-ray Irradiation, Data curation. **M. Ruat:** Investigation, Project administration. **P. Fajardo:** Investigation, Funding acquisition.

**Declaration of competing interest**

The authors declare that they have no known competing financial interests or personal relationships that could have appeared to influence the work reported in this paper.

**Acknowledgements**

This work has been funded by the Spanish Ministry of Science and Innovation (MCIN/AEI/10.13039/501100011033/) and by the European Union's ERDF program "A way of making Europe". Grant references: RTI2018-094906-B-C22 and PID2020-113705RB-C32. Also, it was funded by the European Union's Horizon 2020 Research and Innovation funding program, under Grant Agreement No. 654168 (AIDA-2020), and by the European Synchrotron Radiation Facility (ESRF).